\documentclass[%
 reprint,
nofootinbib,
 amsmath,amssymb,
 aps,
]{revtex4-1}

\usepackage{graphicx}
\usepackage{dcolumn}
\usepackage{bm}
\usepackage{hyperref}

\usepackage{amsmath}
\usepackage{amsfonts}
\usepackage{amssymb}
\usepackage{multirow}
\usepackage{array}
\usepackage{subfigure}
\usepackage{sidecap}

\newcommand{\dslash}{\not\!\partial}

\newcommand{\eps}{\varepsilon}
\newcommand{\vphi}{\varphi}

 \def    \ptj           {\mbox{$p_{T j}$}}
 \def    \etaj           {\mbox{$\eta_{ j}$}}
 \def    \ptl           {\mbox{$p_{T l}$}}
 \def    \etal           {\mbox{$\eta_{ l}$}}
 \def    \gev            {\mbox{$\mathrm{GeV}$}}

\newcommand{\etmiss}{ \not \hskip -4pt E_T}

\begin{document}

\title{Early discovery of top partners and test of the Higgs nature}

\author{Natascia Vignaroli}
 
\affiliation{%
 Department of Physics and Astronomy, Iowa State University, Ames, IA 50011, USA
}%
\affiliation{Department of Physics and Astronomy, Michigan State University, East Lansing, MI 48824, USA.}

\date{\today}

\begin{abstract}
Top partners from a new strong sector can be discovered soon, at the 8 TeV LHC, by analyzing their single production, which exhibits a large enhancement in the cross section compared to 
the analogous productions of bottom partners and exotic quarks. We analyze the subsequent decay of the top partners into a $125$ GeV Higgs. This channel proves to be very promising for both the discovery of top partners and a test of the Higgs sector. \\
For a reference value $\lambda_{\tilde{T}}=3$ of the Higgs coupling to the top partner, we could have a discovery (observation) at the 8 TeV LHC, with $30$ fb$^{-1}$, for top partner masses up to $760$ ($890$) GeV. 
If the LHC and Tevatron excesses near $125$ GeV are really due to a composite Higgs, naturalness arguments demand top partners below $\sim 1$ TeV.  Our results highlight thus that the 8 TeV LHC already has a large sensitivity on probing the composite Higgs hypothesis.\\
The LHC reach is even wider at $\sqrt{s}=14$ TeV.  With $\lambda_{\tilde{T}}=3$ the LHC with $100$ fb$^{-1}$ can observe (at 5$\sigma$) a Higgs from a top partner decay for masses of this latter up to $\simeq 1450$ GeV. In the case the top partner was as light as $\simeq 500$ GeV, the 14 TeV LHC would be sensitive to the measure of the $\lambda_{\tilde{T}}$ coupling in basically the full range $\lambda_{\tilde{T}}>1$ predicted by the theory.
\end{abstract}

\pacs{Valid PACS appear here}
\maketitle


\section{Introduction}
\label{sec:introduction}
It is a compelling possibility that the LHC \cite{lhc-higgs} and Tevatron \cite{tev-higgs} excesses near $125$ GeV could be attributed to a composite Higgs, which is also the pseudo-Goldstone boson arising from some symmetry breaking of a new strong sector \cite{Georgi_Kaplan}. In this case, as several studies \cite{TopMassH} have recently shown,
naturalness arguments demand top partners lighter than $\sim 1$ TeV. In this analysis, we will show that just the top partners, and in particular the partner of the right-handed top, could be discovered soon at the $8$ TeV LHC \footnote{Similar conclusions have also been reached in the context of Little Higgs theories \cite{Perelstein2012}}, by analyzing the channels of their single production. These latter, indeed, exhibit a large enhancement in the cross section, compared to 
the analogous productions of bottom and exotic heavy partners \cite{Mrazek:2009yu}. \\
In this study we will perform a first analysis of the decay to Higgs of a singly-produced top partner ($\tilde{T}$).\\
As pointed out in \cite{Azatov2012, Vignaroli:2012Bs, Santiago2012}, because of the strong interactions among the composite Higgs and the heavy fermions, the analyses of the Higgs production from the decay of a heavy partner represent a promising way also to reveal the composite nature of the Higgs. The analysis of single production, in particular, allows the measurement of the Higgs coupling to the heavy fermions and is thus an important channel to obtain information on the theory behind the EWSB \cite{Vignaroli:2012Bs}. \\
The analysis we will perform is thus promising to both discover the top partners and to understand the Higgs nature.\\
The paper is organized as follows. In Section~\ref{sec:model} we review the effective two-site model that we adopt
to study the phenomenology. 
In Section~\ref{sec:analysis} we study the prospects of observing $\tilde{T}\to ht$ events at the LHC.
We perform a Montecarlo simulation of the signal and the main SM backgrounds and outline a strategy to maximize 
the discovery significance. We discuss the results obtained and draw our conclusions in Section~\ref{sec:conclusions}.


\section{Top partners in a two-site effective theory }
\label{sec:model}

We will work in the framework of an effective theory that reproduces the low-energy limit of a large set of composite Higgs models (CHM) \cite{Agashe:2004rs} and warped extra-dimensional theories with a custodial symmetry in the bulk~\cite{Agashe:2003zs}. 
Specifically, we will adopt a ``two-site'' description \cite{Contino:2006nn}, where two sectors, the weakly-coupled sector of the elementary fields
and the composite sector, that comprises the Higgs, are linearly coupled to each other through mass mixing terms \cite{Kaplan:1991dc}. This leads to a scenario of partial compositeness of the SM; after diagonalization, the 
elementary/composite basis rotates to the mass eigenstate one, made of SM and heavy states that are admixture of elementary and composite modes. \\
In particular, we will refer to the same description of Ref. \cite{Servant}, where composite fermions can be arranged in a 5 of $SO(5)$. This is a minimal model which incorporates the custodial symmetry and the Left-Right parity needed for CHM to pass the EWPT \cite{zbb} and which includes the full set of resonances which are needed to generate the top-quark mass (direct extensions to this description, which include also the resonances needed to give mass to the bottom, and a different spectrum, where composite fermions are in a 10 of $SO(5)$, can be found in \cite{vignaroli_bsGamma}. Additional sets of top partners in the 5 or 10 representations of $SO(5)$ are present in many other works as, for example, \cite{Contino:2006qr,Medina:2007hz}).\\
The two building blocks of the model are the elementary sector and the composite sector. The composite sector has a 
$SO(4) \times U(1)_X$ global symmetry, with $SO(4) \sim SU(2)_L \times SU(2)_R$. The particle content of the elementary sector
is that of the SM without the Higgs, and the $SU(2)_L \times U(1)_Y$ 
elementary fields gauge the corresponding global
invariance of the strong dynamics, with $Y = T_{R}^3 + X$. 
The composite sector comprises   
the composite Higgs
\begin{equation} \label{eq:higgs}
{\cal H} = (\mathbf{2},\mathbf{2})_{0} = 
 \begin{bmatrix} \phi_0^\dagger & \phi^+ \\ - \phi^- & \phi_0 \end{bmatrix} \, ,
\end{equation}
and the following set of vector-like composite fermions:

\begin{align}\label{eq:fermions} 
\begin{split}	
&	\mathcal{Q}=\left[\begin{array}{cc}
	T & T_{5/3} \\ 
	B & T_{2/3} \end{array}\right]=\left(2,2\right)_{2/3} \ , \ \tilde{T}=\left(1,1\right)_{2/3}
 \end{split}
\end{align}

The quantum numbers of the composite fermions and the Higgs under $SU(2)_L \times SU(2)_R \times U(1)_X$ are those
specified in  eqs. (\ref{eq:higgs}), (\ref{eq:fermions}).
The Lagrangian that describes our effective theory
is the following \cite{Georgi_Kaplan} (we work in the gauge-less limit): 
%
\begin{align}  \label{eq:Ltotal}
{\cal L} =&  \, {\cal L}_{elementary} + {\cal L}_{composite} + {\cal L}_{mixing} \\[0.5cm]
\label{eq:Lelem}
 {\cal L}_{elementary} = &\,\bar q_L i\!\dslash\, q_L + \bar t_R i\!\dslash\, t_R \\[0.3cm]
\label{eq:Lcomp}
 {\cal L}_{composite}  = 
&\ \text{Tr}\left\{\bar{\mathcal{Q}}\left(i\dslash -M_{Q*}\right)\mathcal{Q}\right\}\\  \nonumber &+ \text{Tr}\left\{\bar{\tilde{T}}\left(i\dslash-M_{\tilde{T}*}\right)\tilde{T}\right\}
 \nonumber \\
  & + \frac{1}{2} \,\text{Tr}\left\{  \partial_\mu {\cal H}^\dagger \partial^\mu {\cal H} \right\} - V( {\cal H}^\dagger {\cal H}) \nonumber + {\cal L}^{YUK}\\[0.2cm] 
  &
  {\cal L}^{YUK}=Y_{*}\text{Tr}\left\{ \bar{\mathcal{Q}}\mathcal{H}\right\}\tilde{T} \nonumber 
\\[0.3cm]
\begin{split} \label{eq:Lmixing}
{\cal L}_{mixing} =&  -\Delta_{L}\bar{q}_{L}\left(T,B\right)-\Delta_{R}\bar{t}_{R}\tilde{T}+h.c.
\end{split}
\end{align}
%
where $V( {\cal H}^\dagger {\cal H})$ is the Higgs potential. 
$ {\cal L}^{YUK}$ (\ref{eq:Lcomp}) contains the Yukawa interactions among Higgs and composite fermions, with coupling $Y_{*}$. As all the couplings among composites, $Y_{*}$ is assumed to be strong, $1< Y_{*} \ll 4\pi$, where $4\pi$ marks out the non-perturbative regime.
The SM Yukawa coupling of the top arises through the mixings (\ref{eq:Lmixing}) of the elementary fields $t_L$ and $t_R$ to the composite fermions $T$ and $\tilde{T}$, which in turn couple to the Higgs (\ref{eq:Lcomp}). The 
two-site Lagrangian (\ref{eq:Ltotal}) is diagonalized by 
a field rotation from the elementary/composite basis to the mass eigenstate basis~\cite{Contino:2006nn}, that can be conveniently parametrized in terms of the following mixing parameters:
%
\begin{equation}
\tan\vphi_{tR} = \frac{\Delta_{R}}{M_{\tilde{T}*}} \equiv \frac{s_{R}}{c_{R}} , \qquad
\tan\vphi_{L} =\frac{\Delta_{L}}{M_{Q*}}\equiv \frac{s_{L}}{c_{L}} 
\end{equation}
\noindent
Here $\sin\vphi_{tR}$ (shortly indicated as $s_R$) and $\sin\vphi_{L}$ ($s_L$) respectively denote the degree of compositeness of $t_R$ and $(t_L, b_L)$. 
After the diagonalization, the Yukawa Lagrangian 
reads:
 \begin{align}
\begin{split}\label{eq:Lyuk}	
\mathcal{L}^{YUK}=&+Y_{*}s_{L}c_{R}\left(\bar{t}_{L}\phi^{\dag}_{0}\tilde{T}_{R}-\bar{b}_{L}\phi^{-}\tilde{T}_{R}\right)\\
  &-Y_{*}s_{R}\left(\bar{T}_{2/3L}\phi_{0}t_{R}+\bar{T}_{5/3L}\phi^{+}t_{R}\right)\\ 
& -Y_{*}c_{L}s_{R}\left(\bar{T}_{L}\phi^{\dag}_{0}t_{R}-\bar{B}_{L}\phi^{-}t_{R}\right)\\
&+Y_{*}s_{L}s_{R}\left(\bar{t}_{L}\phi^{\dag}_{0}t_{R}-\bar{b}_{L}\phi^{-}t_{R}\right)\\ 
&+\ h.c.+\ \dots\\
\end{split}
\end{align}
where the dots are for terms of interactions among heavy fermions which are not relevant for our analysis. Symbols denoting elementary (composite) fields before the rotation now indicate the SM (heavy) fields.
After the EWSB, the terms in (\ref{eq:Lyuk}) generate the top-quark mass, $m_{t}=\frac{v}{\sqrt{2}}Y_{*}s_{L}s_{R}$. As already discussed, this term comes from the $t_L$ interaction with its heavy-partner $T$ and from the $t_R$ interaction with $\tilde{T}$. \\
The presence of these top partners is a feature of basically all the models which address the hierarchy problem by introducing a new strong sector or a 
warped extra-dimension. $T$ and $\tilde{T}$ play a key role in stabilizing the Higgs. There is thus 
a correlation between the Higgs mass and that of the top partners. As recently shown 
in several studies \cite{TopMassH}, a $\sim 125$ GeV composite Higgs requires the presence of at least one of the top partners with a mass below $1$ TeV. Our analysis will be focused on the search for one of these top partners, the $t_R$-partner $\tilde{T}$. There are no robust constraints from flavor observables that forbid $\tilde{T}$ to be quite lighter than $1$ TeV and, as we will discuss in the next section, the experimental searches put still mild limits on its mass. On the contrary, a robust minimal-flavor-violating bound, of about $1$ TeV, exists on the $T$ mass. Indeed, the $(T,B)$ exchange generates an effective coupling $Wt_Rb_R$ that leads to an important contribution to $b\to s \gamma$ \cite{vignaroli_bsGamma}. In the top composite limit top partners as $T_{2/3}$ or $T_{5/3}$ can become much lighter than the other strong sector's resonances \cite{Contino:2006qr}. They might be lighter than 1 TeV as well, as generally allowed by EWPT and flavor observables \cite{TopMassH,Pomarol:2008bh}.
While $T$ and $\tilde{T}$ are a general prediction of CHM, 
the presence of the exotics $T_{2/3}$ and $T_{5/3}$ is purely related to the custodial symmetry in the strong sector; they could thus be absent in different CHM. 


\subsection{Top partner single production} 

\begin{SCfigure}[]
\includegraphics[width=0.27\textwidth, trim=0cm 0cm 0cm 4cm, clip=false]{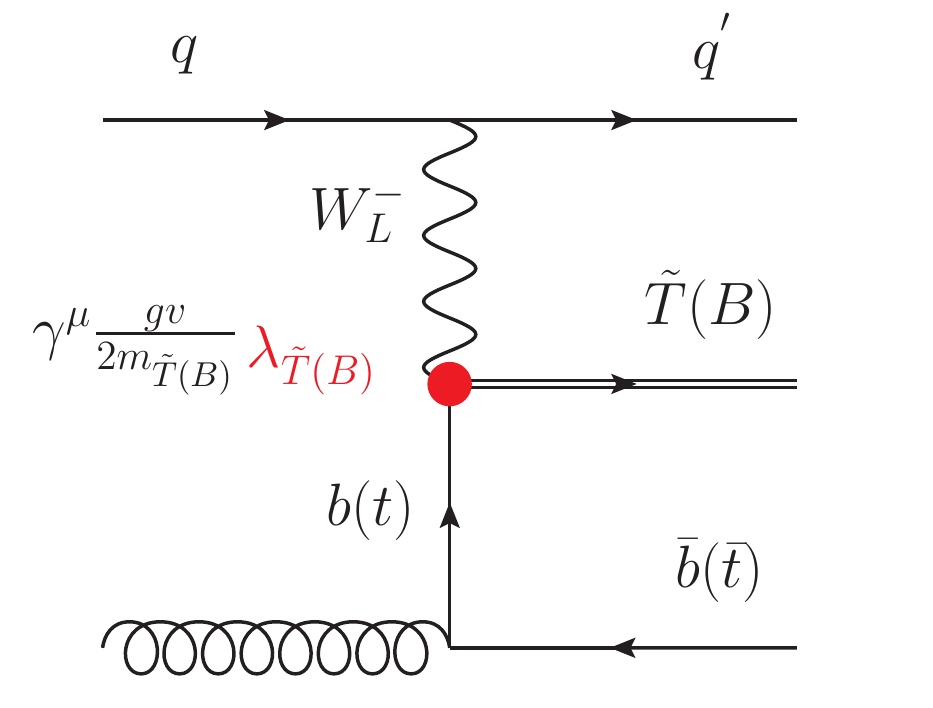}

\caption{ Feynman diagram for the main contribution to the single production of top and bottom partners. 
}
\label{Tsprod_dia}
\end{SCfigure}

\begin{figure}[]
\includegraphics[width=0.45\textwidth]{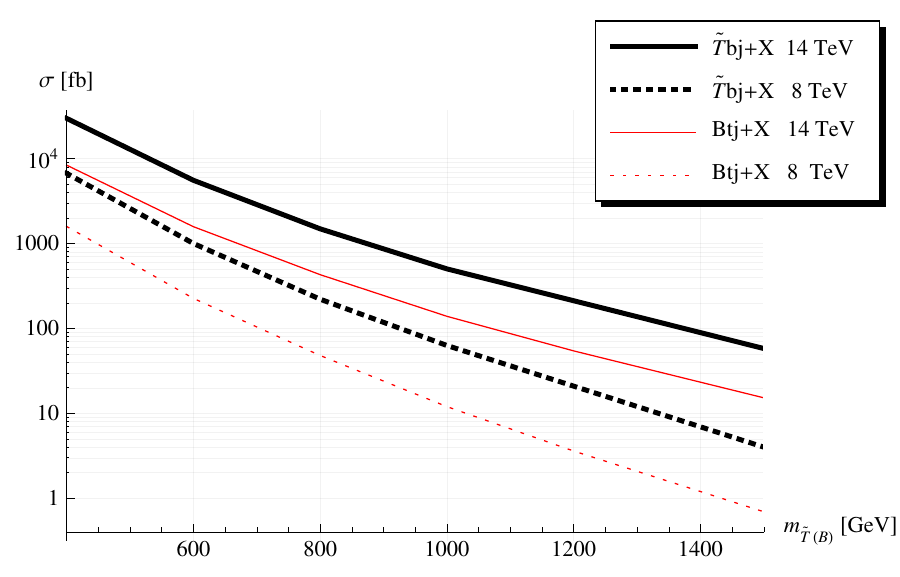}
\caption{
Cross section values for the single production of top (black curves) and bottom (red curves) partners. We summed over the single production of $\tilde{T}$($B$) and $\bar{\tilde{T}}$($\bar{B}$) at the LHC with $\sqrt{s}=14$ TeV (continuous lines) and $\sqrt{s}=8$ TeV (dotted lines). We set the $\lambda_{\tilde{T}}$ and the $\lambda_{B}$ coupling to the reference value $\lambda=3$; cross sections scale with $\lambda$ as $\lambda^2$.}
\label{Tsprod_xsec}
\end{figure}

 Heavy fermionic resonances can be singly-produced through their interactions with longitudinal electro-weak bosons and SM quarks (see Fig. \ref{Tsprod_dia}). These interactions originate from the Yukawa terms in (\ref{eq:Lyuk}), after the EWSB, as a consequence of the Yukawa couplings among composites (\ref{eq:Lcomp}) and the mixing between composite and elementary fermions (\ref{eq:Lmixing}). 
 Top partner single production can also occur at LO in QCD coupling (see Ref. \cite{Perelstein2012} for a recent analysis) through the exchange of a $W$ in the t-channel, $qb\to q^{'}\tilde{T}$ (which is the dominant LO contribution at the LHC), or in the s-channel, $qq^{'}\to b\tilde{T}$. Notice that the associated $W$ production, $qg\to W\tilde{T}$, is suppressed in CHM, because $gt\tilde{T}$ couplings are forbidden by gauge invariance. 
 We choose to focus our analysis on the NLO production in Fig. \ref{Tsprod_dia} because its cross section is comparable with the LO one, since the process requires only initial light quarks and gluons, and, most importantly, because the presence of the extra $b$ quark in the signal leads to a very distinctive topology that gives an advantage to overcome the background \footnote{The extra $b$ quark in the signal is emitted, for the almost total part of the events, in the central region $|\eta_b|<2.5$ (see Fig. \ref{MRj_distribution}).  Moreover, the set of cuts we will impose tends to select a very energetic final state and, as a consequence, a $b$ quark at high $p_T$. These aspects ensure the reliability of QCD perturbation theory, avoiding the large log's in the region where $b$ is collinear with the incoming proton.}.
 Fig. \ref{Tsprod_xsec} shows the cross sections for the NLO single production of top ($\tilde{T}$) and bottom ($B$) partners at the LHC with $\sqrt{s}=14$ TeV and $\sqrt{s}=8$ TeV.  
The $B$ heavy-bottom, as well the exotic $T_{5/3}$, can only be singly-produced from their interaction with the longitudinal $W$ and the SM top; the couplings of this interaction, $\lambda_B=Y_{*}c_L s_R$, $\lambda_{T5/3}=Y_{*}s_R$, can be directly read from the Yukawa Lagrangian (\ref{eq:Lyuk}). The $B$ (and the $T_{5/3}$) production is thus accompanied by the exchange of a top. On the other hand, the $\tilde{T}$ single production can proceed through the interaction with a $W_L$ and a bottom, with coupling (\ref{eq:Lyuk}) 
\begin{equation}\label{eq:lambda}
\lambda_{\tilde{T}}=Y_{*}s_L c_R \ . 
\end{equation}
Because of the exchange of the bottom instead of the top, the $\tilde{T}$ single production has a much higher (about 4-5 times) cross section than that of the $B$ (and $T_{5/3}$) single production, as clearly shown by Fig. \ref{Tsprod_xsec}. 
Top partners could thus represent, also considering the naturalness argument, the first type of heavy-fermions to be observed at the LHC, through their single production.   \\
 In Fig. \ref{Tsprod_xsec} we consider both the heavy fermions charges. Notice that, due to the different content of the up and down partons in the proton, the cross section for the single production of $\tilde{T}$ is roughly 2 times that for $\bar{\tilde{T}}$. 
 In our analysis we will exploit mainly kinematic cuts and we will not make use of this charge asymmetry. Anyway it could represent a promising variable to discriminate between signal and background. 
 In this analysis we will focus on the $\tilde{T}$ single production and on its subsequent decay into the composite Higgs, $pp\to (\tilde{T}\to ht)b+X$. We will also consider the leptonic decay of the top and the decay of the Higgs into a $b\bar{b}$ pair, with the same branching ratio of that of a SM Higgs of $125$ GeV \footnote{In the case the composite Higgs is an approximate Goldstone boson (a scenario which is preferred by EWPT, flavor and naturalness arguments), its branching ratios are modified in respect to the SM. The modification mainly depends on the value of the Higgs decay constant $f$. However, except for particular values of $f$, $f^2 \sim 2 v^2$, for which the Higgs might become fermiophobic, the $H \to b\bar{b}$ branching ratio is very close to the SM one (see Ref. \cite{Contino:2010mh} and in particular Fig. 2 in it for more details).}.  Notice that the same process, with the exchange of the intermediate $W_L$ and the bottom, does not occur for the $T$ and the $T_{2/3}$ top partners, that, at LO, do not interact with $W_L$ and $b$ (see eq. (\ref{eq:Lyuk})). Their single production, as for the $B$ and the $T_{5/3}$, is accompanied by the exchange of a top and is thus lower than that of $\tilde{T}$. The interactions beyond the SM which are relevant for our analysis can be restricted to the following terms: 
 
 \begin{align}  \label{eq:LTs}
\begin{split} 
 {\cal L}_{\tilde{T},h}  =
  & \bar{\tilde{T}}\left(i\dslash -m_{\tilde{T}}\right)\tilde{T} + \frac{1}{2} \,\text{Tr}\left\{  \partial_\mu {\cal H}^\dagger \partial^\mu {\cal H} \right\} - V( {\cal H}^\dagger {\cal H}) \\[0.2cm]
  & +Y_{*}s_{L}c_{R}\left(\bar{t}_{L}\phi^{\dag}_{0}\tilde{T}_{R}-\bar{b}_{L}\phi^{-}\tilde{T}_{R}\right) + h.c.
\end{split} 
\end{align}

The $\tilde{T}$ branching ratios (BR) are essentially fixed by the equivalence theorem to be $BR(\tilde{T}\to W_L b)\simeq 0.50$,  
$BR(\tilde{T}\to Z_L t)\simeq BR(\tilde{T}\to ht)\simeq 0.25$. The rates for the $\tilde{T}$ decays and single production depend quadratically on the coupling (\ref{eq:lambda}), $\lambda_{\tilde{T}}= Y_{*} s_L c_{R}$, which can be directly read from the Lagrangian (\ref{eq:LTs}). We remind that $Y_{*}$ represents the Yukawa coupling among composite states (see eq. (\ref{eq:Lcomp})) and it is assumed to be large. 

Despite we are referring to a specific composite Higgs model, our analysis could be easily extended to other scenarios, considering that it depends on just two parameters (once we fix the Higgs mass and the $h\to b\bar{b}$ branching ratio): 
$\lambda_{\tilde{T}} \ ,\ m_{\tilde{T}} $ .

As already said, a $\tilde{T}$ top partner, which can be singly-produced, is present in a wide class of models with a new strong sector, a warped extra-dimension or in Little Higgs theories (see, for example, the studies in \cite{Little-Higgs-analyses} in this latter context). Moreover, if we consider different representations for composite fermions, other top partners, which can be singly-produced as the $\tilde{T}$, could also be present.  An example of this is the $\tilde{T}^{'}$, which appears in a $10$ of $SO(5)$ \cite{vignaroli_bsGamma, Vignaroli:2012Bs, Contino:2006qr, Medina:2007hz}, and which is expected to be even lighter than the $\tilde{T}$.\\
At the end we will analyze the LHC sensitivity on the $(\tilde{T}\to ht)b+jets$ channel in the $(m_{\tilde{T}}, \lambda_{\tilde{T}})$ parameter space. We now fix $\lambda_{\tilde{T}}=3$ and we consider several $\tilde{T}$ mass values. $\lambda_{\tilde{T}}=3$ could be realized, for example, if we have a Yukawa coupling $Y_{*}\sim 3$ (which is a very plausible value, given the assumption of large couplings among composites, $1<Y_*\ll4\pi$) 
and a left-handed top with a large degree of compositeness, $s_L\sim 1$ (which implies $c_R\sim 1$) \footnote{The scenario of a fully composite left-handed top, $s_L = 1$, could account for the heaviness of the top quark and is allowed by EWPT \cite{Pomarol:2008bh}.}.  We consider a quite wide $\tilde{T}$ mass range, $m_{\tilde{T}}\ge 400$ GeV. The strongest lower bound on $m_{\tilde{T}}$, that comes from the LHC data at $\sqrt{s}=7$ TeV, is indeed of about $420$ GeV. We derive this constraint from the study in \cite{mTsBound}, which presents a search for pair production of top-prime's decaying predominantly into $Wb$, by considering a value $BR(\tilde{T}\to Wb)=0.5$ instead of $BR(\tilde{T}\to Wb)=1$.    \\
In the range of masses we will consider and for the reference value $\lambda_{\tilde{T}}=3$ we have 
\begin{equation} 
BR(\tilde{T}\to ht)\simeq 0.25  \qquad \Gamma(\tilde{T})/m_{\tilde{T}}\simeq 0.17
\end{equation}

\section{Analysis}
\label{sec:analysis}

In this section we discuss the prospects of observing the process $pp\to (\tilde{T}\to (h\to bb)t)b+X$ at the LHC. We consider the leptonic decay of the top. The physical final state is thus:
\begin{equation} \label{eq:finalstate}
pp \to l^\pm \! + n\, jets \, +  \not\!\! E_T\ .
\end{equation}
We will present a simple parton-level analysis aimed at assessing the LHC discovery reach. 
We consider two center-of-mass energies: $\sqrt{s}=8\,$TeV,
the energy of the current phase of data taking, and $\sqrt{s}=14\,$TeV, the design energy that will
be reached in the next phase of operation of the LHC. Our selection strategy 
does not depend, however, on the value of the collider energy. This is because we will apply a set of cuts which exploit the peculiar kinematics of the signal, and a change in the collider energy mainly implies a rescaling of the production cross sections of signal and background via the parton luminosities, without affecting the kinematic distributions.

\subsection{Montecarlo simulation of signal and background}

We simulate the signal by using MadGraph v4~\cite{MG-ME}, after implementing the two-site model with Feynrules \cite{Feynrules}, 
while for the background we make use of both MadGraph and ALPGEN~\cite{Mangano:2002ea}. 
In our parton-level analysis jets are identified 
with the quarks and gluons from the hard scattering.
If two quarks or gluons are closer than the separation $\Delta R =0.4$, they are merged into a single jet whose four-momentum is the
vectorial sum of the original momenta.
We require that the jets and the leptons satisfy 
the following set of acceptance and isolation cuts:
\begin{equation}
\begin{aligned}
\ptj &\geq 30\, \gev  & \quad  | \etaj | &\leq 5    & \quad 
 \Delta R_{jj} &\geq 0.4 \\[0.2cm]
\ptl &\geq 20\, \gev & \quad  | \etal | &\leq 2.5  & \quad 
 \Delta R_{jl} &\geq 0.4 \, .
\end{aligned}
\label{eq:acceptance}
\end{equation}
Here $\ptj$ ($\ptl$) and $\etaj$ ($\etal$) are respectively 
the jet (lepton) transverse momentum and pseudorapidity, and 
$\Delta R_{jj}$, $\Delta R_{jl}$ denote the jet-jet and jet-lepton separations.

Detector effects are roughly accounted for by performing a simple Gaussian smearing on the 
jet energy and momentum absolute value with $\Delta E/E = 100\%/\sqrt{E/\text{GeV}}$, and on the jet momentum direction using
an angle resolution $\Delta\phi =0.05$ radians and $\Delta\eta = 0.04$. Moreover, the missing energy $\etmiss$ of each event has been computed
by including a Gaussian resolution $\sigma(\etmiss) = a \cdot \sqrt{\sum_i E^i_T/\gev}$, where $\sum_i E^i_T$ is the scalar sum of the transverse
energies of all the reconstructed objects (electrons, muons and jets). We choose $a=0.49$~\footnote{This numerical value, as well as the $b$-tagging efficiency and rejection rate and the resolution parameters considered in the jet smearing, have been chosen according to the performance of the ATLAS detector~\cite{Aad:2008zzm}.} .
After applying the acceptance and isolation cuts (\ref{eq:acceptance}) to the signal, we find a fraction of signal events with four reconstructed jets in the final state (where jet is either a light jet or a $b$-jet) of about 0.4. The signal events with $4j$ are mainly constituted by events where one of the final five jets is soft, with $\ptj<30$ GeV, and does not pass the cuts in (\ref{eq:acceptance}) \footnote{The fraction of signal events where one jet is a `fat' jet, resulting from the merging of the $b\bar{b}$ pair from the Higgs, increases with larger $\tilde{T}$ masses, because the Higgs is more boosted in this case, but is not so relevant (it is about 0.07 for a $1$ TeV $\tilde{T}$); sophisticated technique of tagging boosted object could not be useful in this case.}. 
Because this fraction is significant and we would want to preserve it, we will select events with at least four jets passing the cuts in (\ref{eq:acceptance}) and exactly one lepton from the leptonic decay of the top quark. We further require the $b$-tagging of at least two $b$-jets. %
\begin{equation} \label{eq:evsel}
pp \to l^\pm \! + n\, jets \, +  \not\!\! E_T\, , \ \ n\geq 4 \, , \ \ \ \text{At least 2 $b$-tag}
\end{equation}
where all objects must satisfy the acceptance and isolation cuts of eq.(\ref{eq:acceptance}). 

The largest SM background after the event selection of eq.(\ref{eq:evsel}) is the irreducible background $WWbb+jets$, 
which includes the resonant sub-processes $Wtb+jets$ (single top) and $t\bar t +jets$.
The latter, in particular, gives the largest contribution.
Another background which will turn out to be relevant after imposing our full set of kinematic cuts is $Wbb+jets$. The background $W+jets$, where at least two of the light jets are mistagged as $b$-jets, will turn out to be negligible after imposing our set of cuts. \\
We estimate the efficiency of tagging at least two $b$-jets by considering binomial distributions with $\eps_b = 0.6$ being the efficiency of the $b$-tag, and $\zeta_b = 0.01$ the probability of mistagging a light jet as a $b$-jet; we also require the jet to be in the central region ($|\eta_j|<2.5$) as a condition for the $b$-tag. We find efficiencies of about 0.81 for the signal events (which have typically four central $b$ quarks), of about 0.35 for the $WWbb+jets$ and the $Wbb+jets$ backgrounds (with typically two central $b$ quarks) and of about $4\cdot 10^{-4}$ for the $W+jets$ background.

We have simulated the $WWbb$ events by using MadGraph, while the other backgrounds are generated with ALPGEN ~\footnote{
The factorization and renormalization scales have been set to be equal and chosen as follows:
$Q= m_{\tilde{T}}/4$ for the signal; $Q = \sqrt{m_W^2 + \sum_j p_{T j}^2}$ for $WWbb+jets$; 
$Q = \sqrt{m_W^2 +  p_{T W}^2}$ for $Wbb+jets$ and $W+jets$.
}.
For simplicity, in our analysis we include all the samples with increasing multiplicity of light jets in the final state. 
This is a redundant procedure which could lead to a double counting of kinematic configurations.
A correct procedure would be resumming soft and collinear emissions by means of a parton shower,
and adopting some matching technique to avoid double counting. However, retaining all the $W+ n\, jets$ samples, we expect to obtain a conservative estimate
of the background. Moreover some of the cuts we will impose tend to suppress
the events with larger number of jets and thus to reduce the amount of double counting.

%
%

\subsection{Reconstruction of the top partner and of the Higgs resonances}\label{sec:reco}

We are focusing our analysis on the channel $pp\to (\tilde{T}\to (h\to bb)t)b+X$. 
The physical final state is that in eq. (\ref{eq:finalstate}).\\
We have $n\geq 4$ jets in the final state. Two of these jets should not come from $\tilde{T}$, as one is the light jet from the initial parton that emits the intermediate $W$, the other is the $b$-jet coming from the initial gluon (Fig. \ref{Tsprod_dia}). In order to reconstruct the $\tilde{T}$ resonances and the intermediate final state $\tilde{T}jb$ (see Fig. \ref{Tsprod_dia}), we need to tag these two jets.\\
As a first step we can easily tag the light-jet of the $\tilde{T}jb$ final state. 
We can do this by considering that this jet tends to be emitted at very high rapidity. 
As also discussed in \cite{Mrazek:2009yu, Vignaroli:2012Bs} and first found in \cite{Willenbrock}, the intermediate $W$ tends to carry only a small fraction of the initial parton energy, in order to maximize its propagator. At the same time, it must have enough energy to produce the heavy top. Thus, the quark in the final state that originates from the parton emitting the $W$ has a high energy and a small transverse momentum (we find a ratio of about ten between the quark energy and the quark transverse momentum). This results in a final light jet with high rapidity, $|\eta|\gtrsim$ 2. This is a peculiarity of the topology of the signal that we exploit to reconstruct the $\tilde{T}$ resonance and that we will also further exploit to discriminate between the signal and the background.
We check from the Montecarlo simulation that the light-jet in the signal represents the jet with the highest rapidity in about the $70-80 \%$ of the cases. The fraction is 0.71 for $m_{\tilde{T}}=0.4$ TeV and grows up to 0.78 for a $\tilde{T}$ of $1$ TeV. We thus tag the light-jet in $\tilde{T}jb$, by assuming that it coincides with the jet with the highest rapidity.\\
%
We thus proceed to tag the top and its decay products.  
The procedure we adopt requires the reconstruction of the momentum of the neutrino.
The transverse momentum of the neutrino can be reconstructed from the transverse missing momentum;
this latter can be estimated, considering a $p^{TOT}_T=0$ hypothesis, as $p^{miss}_T=-\sum p_T$, where $\sum p_T$ is the sum over the $p_T$ of all the detected final states.
Once we have estimated the neutrino transverse momentum, we can derive the neutrino longitudinal momentum, $p_z$, by requiring that the
neutrino and the lepton reconstruct an on-mass-shell $W$, $M_{l\nu} = 80.4$ GeV. The condition
\begin{equation}
 (E^{l}+E^{\nu})^{2}-(p^{l}_x+p^{\nu}_x)^{2}-(p^{l}_y+p^{\nu}_y)^{2}-(p^{l}_z+p^{\nu}_z)^{2}=M^2_W
\label{neutrino}
\end{equation}
 gives two solutions for $p^{\nu}_z$. 
 We find that in the $\simeq 20 \%$ of the events, both for the signal and the background, eq. (\ref{neutrino}) has imaginary solutions 
(this corresponds to the case of a quite off-shell leptonically decayed $W$). In this case we decide to reject the event. 
Our neutrino reconstruction procedure has, therefore, an efficiency of about the $80\%$. 
Once we have reconstructed the momentum of the neutrino, we want to reconstruct the top which is in our signal and to tag the jet associated to its decay.
 To do this, we first reconstruct the leptonically decayed $W$ and then we consider all the possible $Wj$ combinations between the $W$ and the jets in the final state, with the exception of the previously tagged light-jet. 
In each event we have two $W\to l\nu$ candidates, one for each of the two solutions of the neutrino longitudinal momentum, eq.(\ref{neutrino}).
The $Wj$ pair that gives the $M_{Wj}$ invariant mass closest to the top mass, $m_t=174$ GeV, is selected as the pair coming from the decay of the top. Notice that this procedure allows, as a bonus, to fully reconstruct the neutrino.
The top 4-momentum is then reconstructed by summing on the 4-momentum of the $W$ and of the jet that form the selected pair. 
We require the invariant mass of the reconstructed top to be in a range $[160$ GeV, $190$ GeV$]$;
this condition reduces the $Wbb+jets$ background, which does not contain a real top; on the contrary, it does not affect the $WWbb$ background and, most importantly, the signal.\\
At this point the not-yet-tagged final jets are those coming from the Higgs and the $b$-jet of the $\tilde{T}jb$ final state. \\
This latter, similarly to the tagged light-jet, tends to be emitted at a much higher rapidity than those of the jets from the Higgs. We thus select the jet with the highest rapidity, among all the not-yet-tagged final jets, as the $b$-jet of the $\tilde{T}jb$ final state. \\
The Higgs can be thus easily reconstructed by considering all of the remaining jets as its constituents. 
The heavy-top is finally reconstructed by summing on the 4-momentum of the reconstructed Higgs and top particles.\\
The reconstruction of the $\tilde{T}$ and Higgs resonance is crucial for the discovery of such particles and to obtain an estimate value of their masses. The reconstruction of the intermediate final state $\tilde{T}jb$ is also very useful to design a strategy for the reduction of the background.

\subsection{Event selection}

In Table \ref{tab:cutflow1} we report the value of the cross section for the signal and the 
main SM backgrounds after the selection (\ref{eq:evsel}) based on the acceptance and isolation cuts of eq.(\ref{eq:acceptance}) and the $b$-tag efficiencies, 
and after the neutrino and top reconstruction, 
for $\sqrt{s} = 8\,$TeV and $\sqrt{s} = 14\,$TeV.

%
\begin{table}
\begin{center}
{\small
\begin{tabular}{|c|cc|cc|}
\multicolumn{1}{c}{} & \multicolumn{2}{c}{\textsf{LHC 8 TeV}} &   \multicolumn{2}{c}{\textsf{LHC 14 TeV}}\\[0.1cm]
\hline
 & \multicolumn{1}{c}{\textsf{acceptance}} & \multicolumn{1}{c|}{top reco} & \multicolumn{1}{c}{\textsf{acceptance}} & \multicolumn{1}{c|}{top reco} \\[0.1cm]
 & & & & \\[-0.3cm]
$m_{\tilde{T}}=0.4\,$TeV & 97.4 & 71.5 & 434 & 317 \\[0.25cm]
$m_{\tilde{T}}=0.6\,$TeV & 16.8 & 11.8 & 94.5 & 65.4 \\[0.25cm]
$m_{\tilde{T}}=0.8\,$TeV & 3.89 & 2.65 & 26.5 & 17.8 \\[0.25cm]
$m_{\tilde{T}}=1.0\,$TeV & 1.11 & 0.735 & 8.98 & 5.91 \\[0.25cm]
$m_{\tilde{T}}=1.5\,$TeV & & & 1.02 & 0.806 \\[0.35cm]
  $WWbb$            & 3510 & 2490  & 16700 & 11000\\[0.25cm]
  $WWbbj$            & 2160 & 1590 &  10600 & 7790\\[0.25cm]
  $WWbbjj$            & 800 & 572 & 4640 & 3210\\[0.25cm]
  $Wbbjj$                & 137  & 63.5 & 573 & 247 \\[0.25cm]
  $Wbb3j$                &  52.9  & 26.3 & 324 & 150 \\[0.25cm]
  $W4j$              &  11.2  &  5.19 & 38.4 & 16.5\\[0.25cm]
  $W5j$             &  4.42  &  2.23 & 18.7 & 8.61 \\[0.35cm]
  Total                  & & & &\\
  background       & 6680 &  4750 & 33000 & 22400  \\[0.15cm]
\hline
\end{tabular}
}
\caption{
\label{tab:cutflow1}
\small 
Cross sections, in fb, for the signal (with $\lambda_{\tilde{T}}=3$) and the main backgrounds after the selection (\ref{eq:evsel}) based on the acceptance cuts of eq. (\ref{eq:acceptance}) and the $b$-tag efficiencies and
after the reconstruction of the neutrino and of the top quark, for $\sqrt{s} = 8\,$TeV and $\sqrt{s} = 14\,$TeV.
}
\end{center}
\end{table}

\begin{figure}[]
\begin{center}
\includegraphics[width=0.235\textwidth,clip,angle=0]{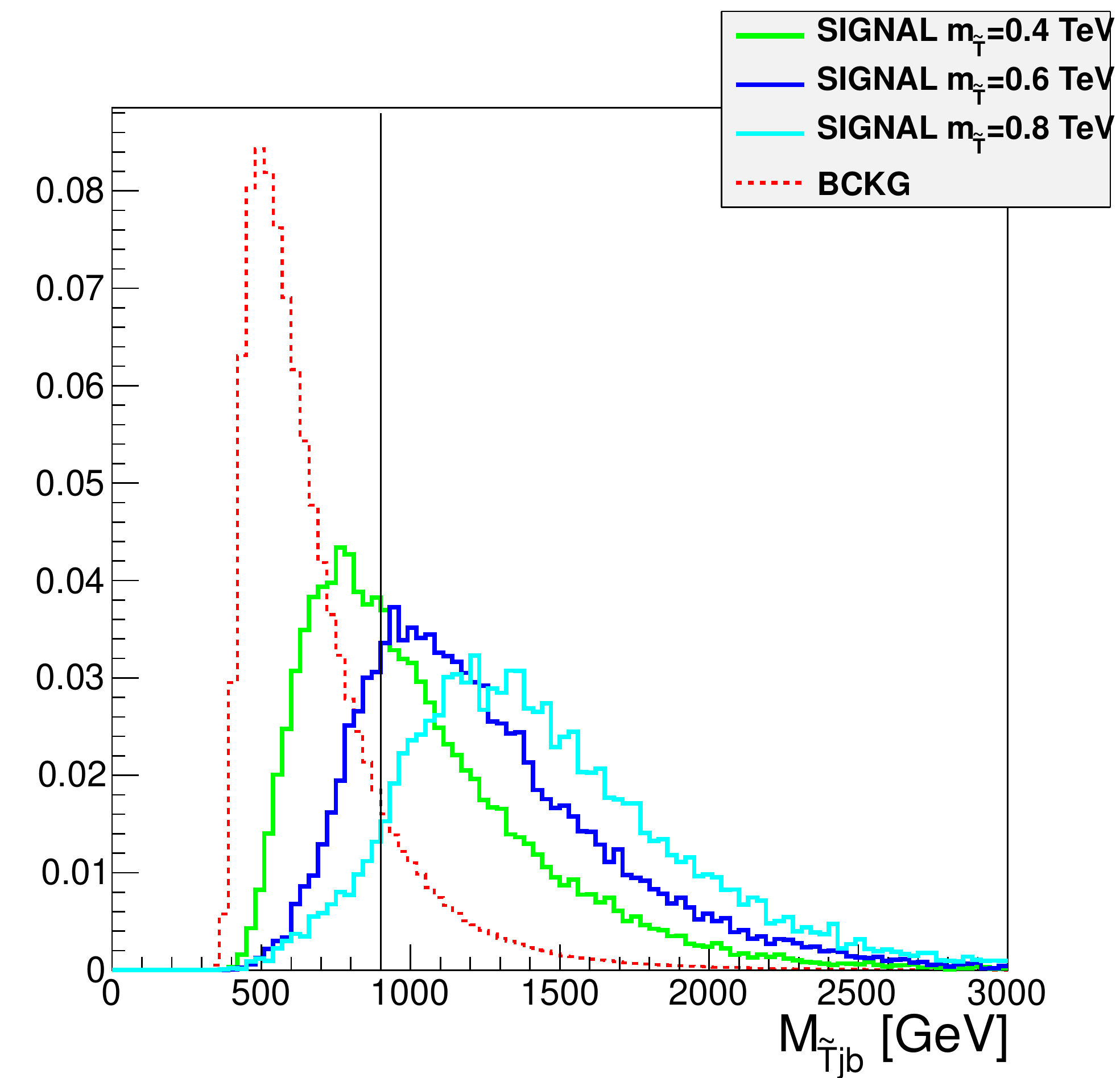}
\includegraphics[width=0.235\textwidth,clip,angle=0]{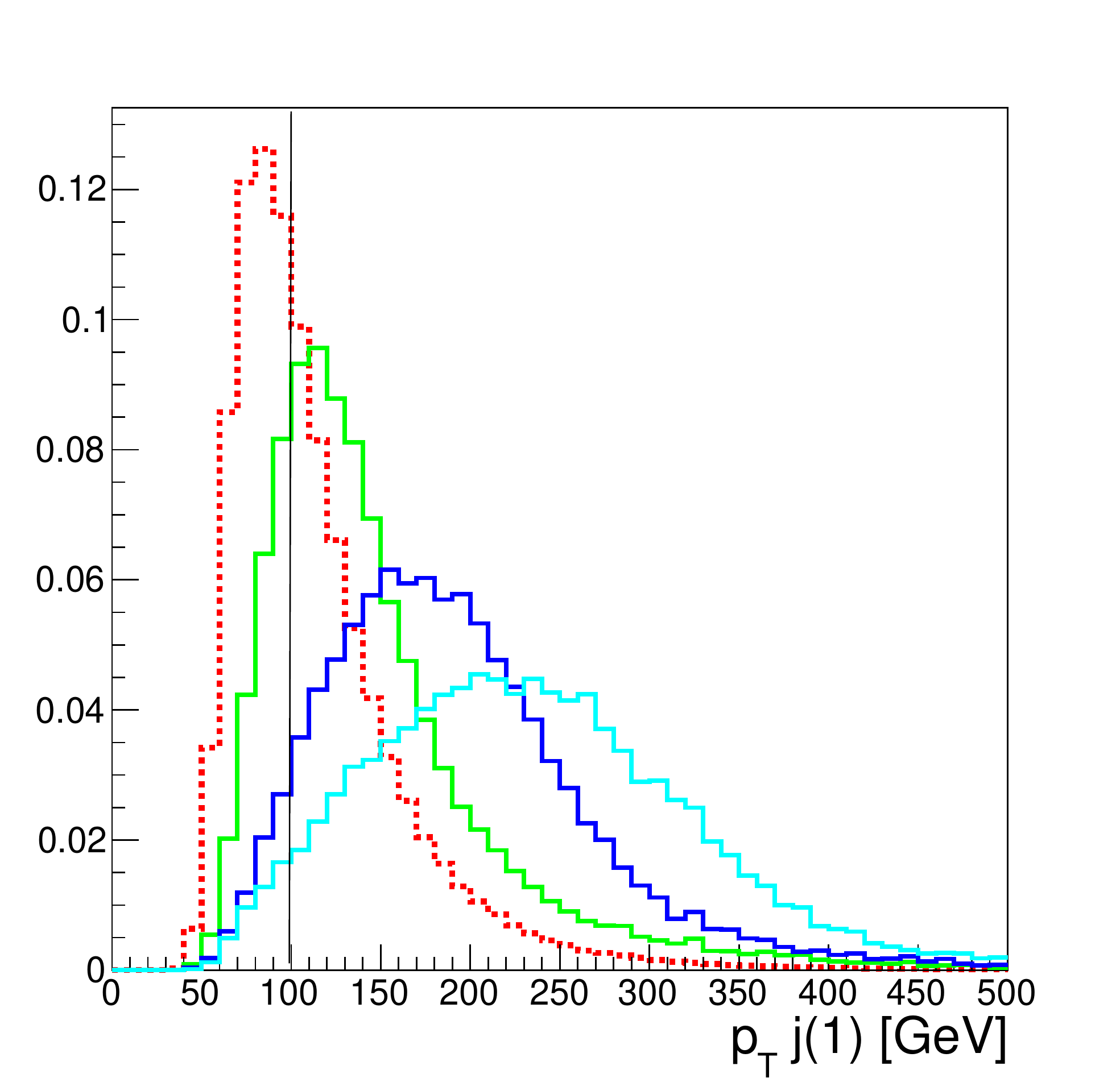}
\\[0.05cm]
\includegraphics[width=0.235\textwidth,clip,angle=0]{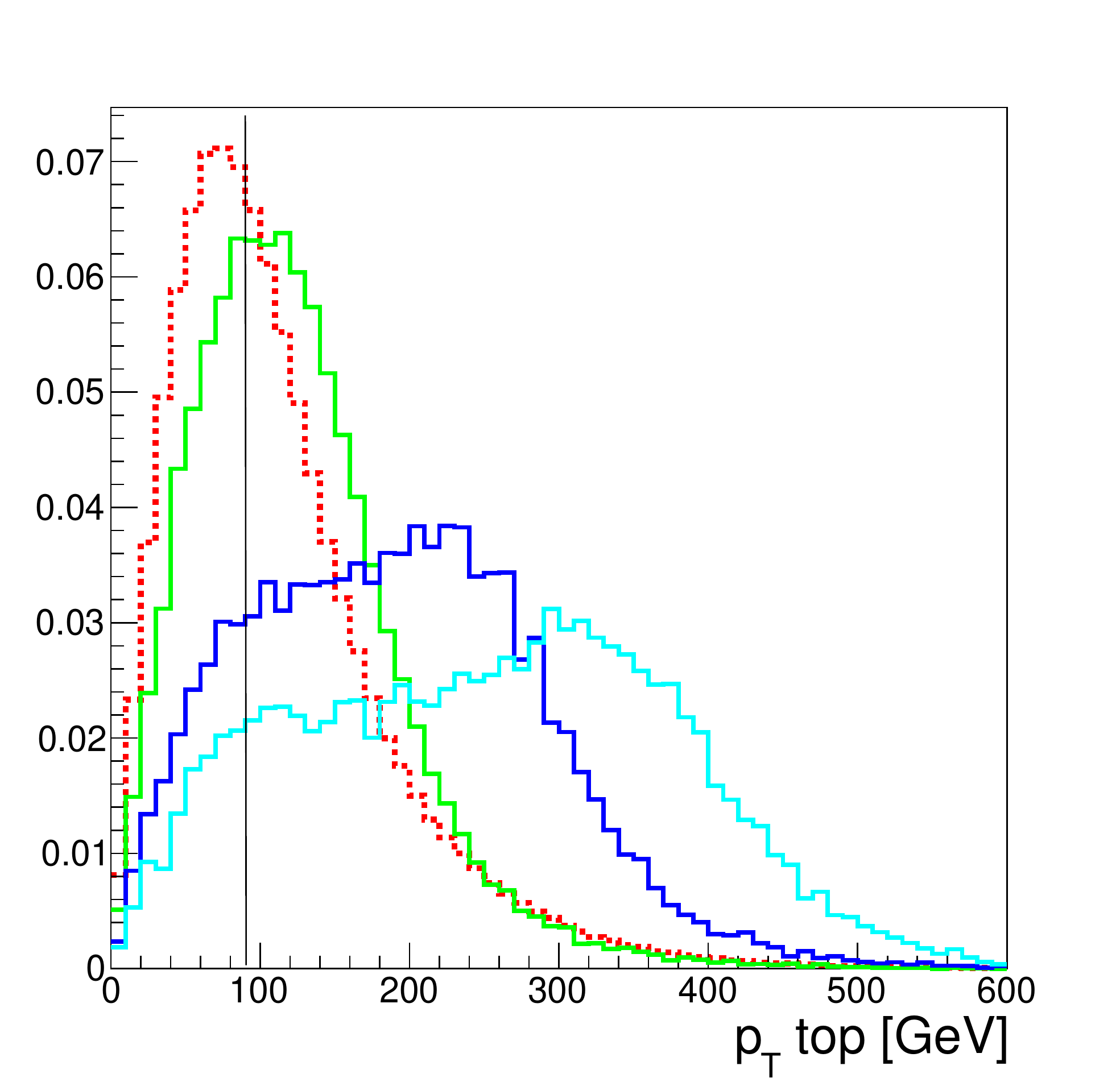}
\includegraphics[width=0.235\textwidth,clip,angle=0]{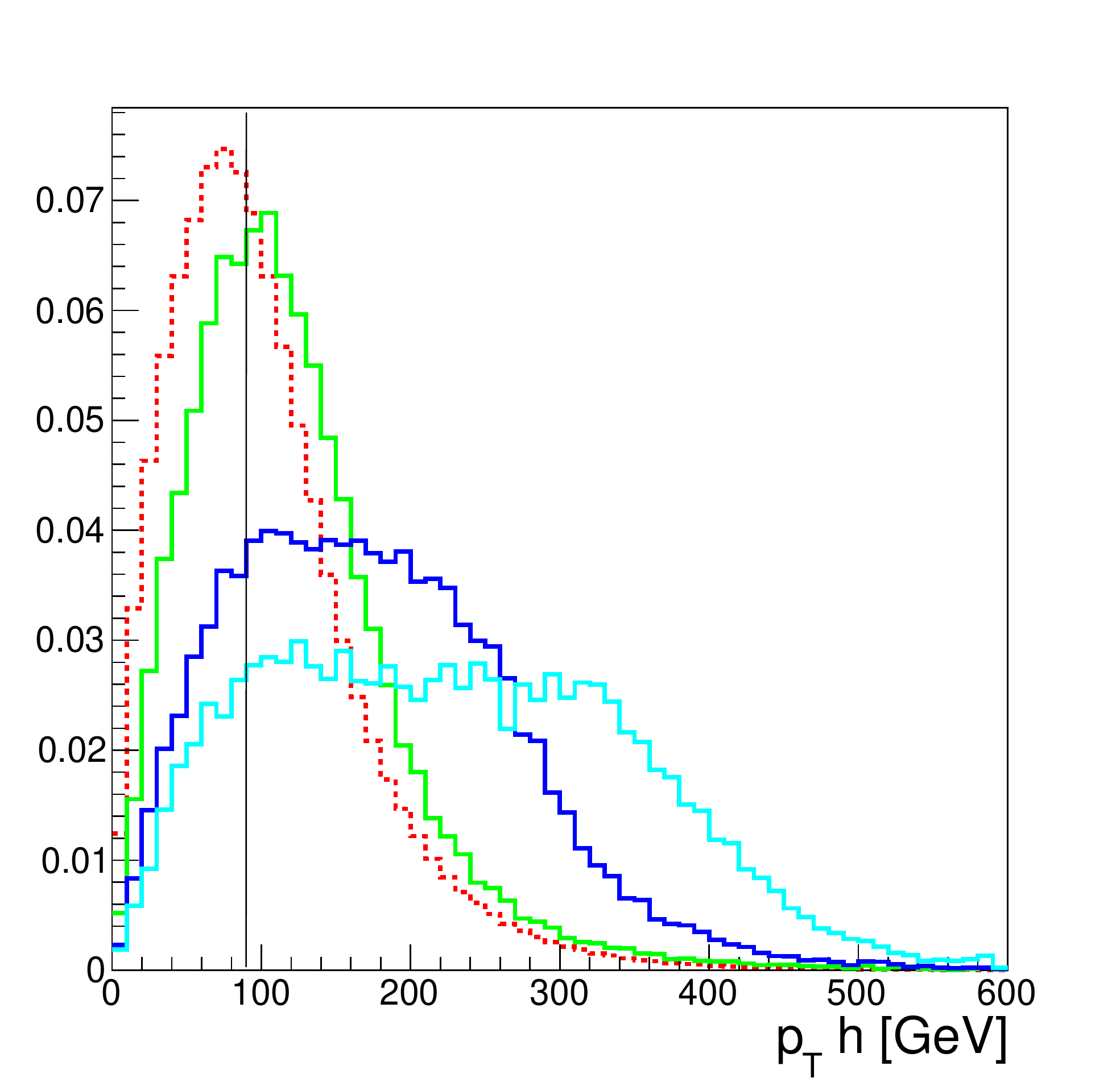}
\caption[]{
\label{pT_distribution}
\small
Differential distributions after the neutrino and  top
reconstruction, 
for $\sqrt{s} =8\,$TeV.
Upper left plot: invariant mass of the system $\tilde{T}jb$
. Upper right plot: $p_T$ of the hardest jet. Lower left plot: $p_T$ of the reconstructed top.
Lower right plot: $p_T$ of the reconstructed Higgs.
The continuous lines show the signal at different $\tilde{T}$ mass values, the dashed (red) line
shows the total background.
All the curves have been normalized to unit~area. The vertical lines indicate the values of the optimized cuts (\ref{eq:optimized}).
}
\end{center}
\end{figure}

\begin{figure}[tbp]
\begin{center}
\includegraphics[width=0.235\textwidth,clip,angle=0]{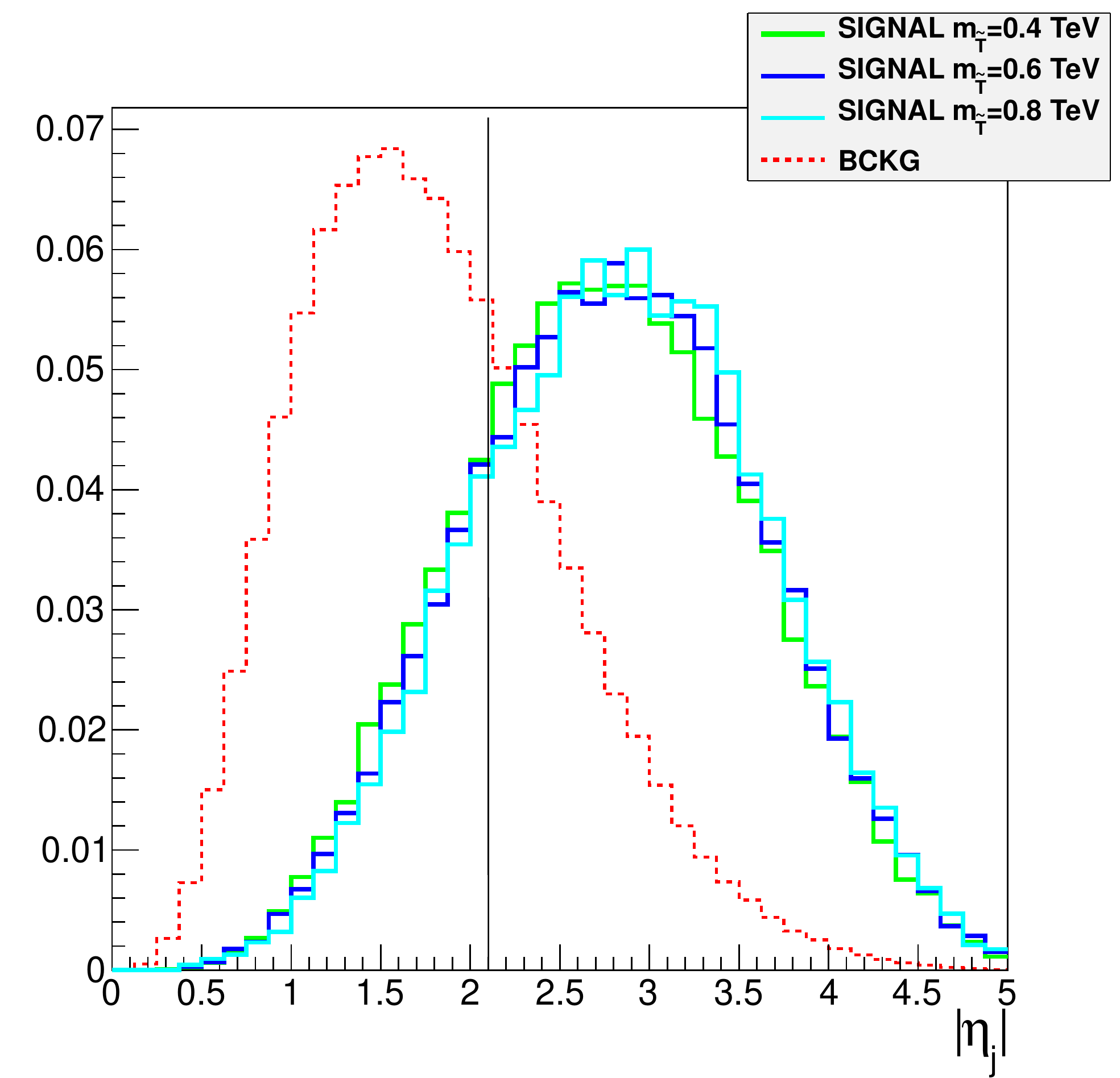}
\includegraphics[width=0.235\textwidth,clip,angle=0]{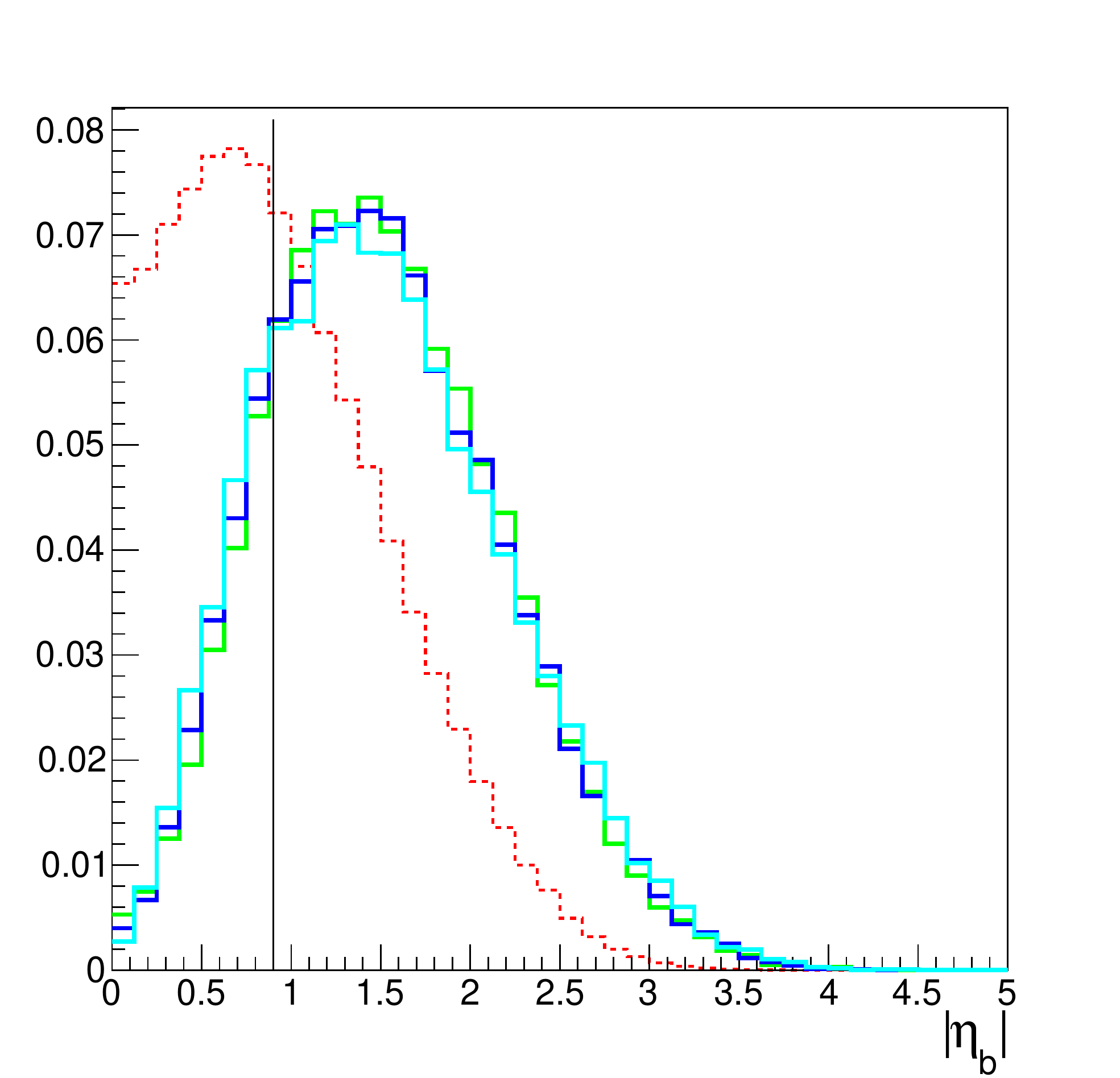}
\caption[]{
\label{MRj_distribution}
\small
Differential distribution of the rapidity of the tagged light-jet (left plot) and of the tagged $b$-jet (right plot), after the neutrino and  top
reconstruction, 
for $\sqrt{s} =8\,$TeV. The vertical lines indicate the values of the optimized cuts (\ref{eq:optimized}).
%
}
\end{center}
\end{figure}

\begin{figure}[tbp]
\begin{center}
\includegraphics[width=0.235\textwidth,clip,angle=0]{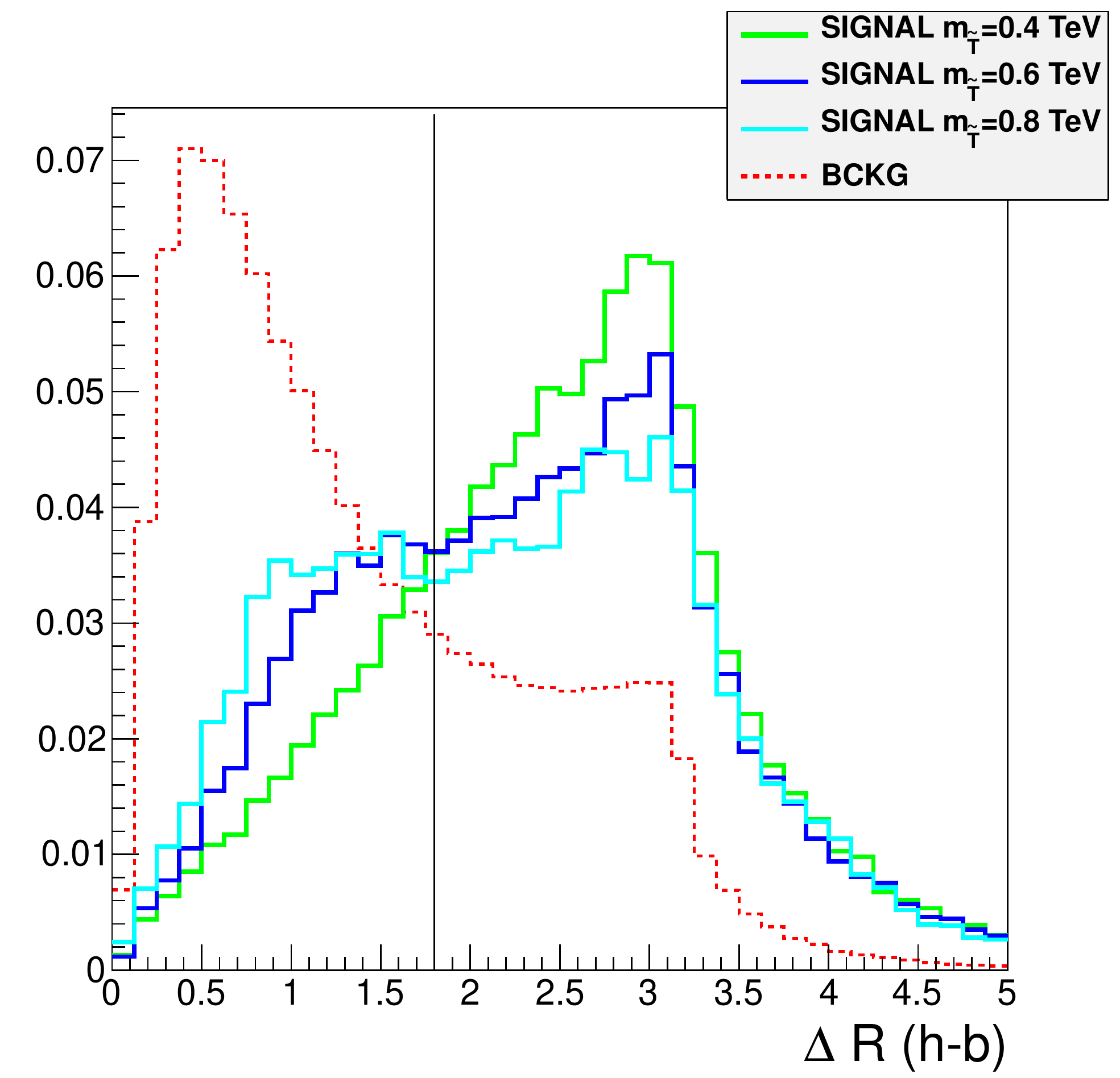}
\includegraphics[width=0.235\textwidth,clip,angle=0]{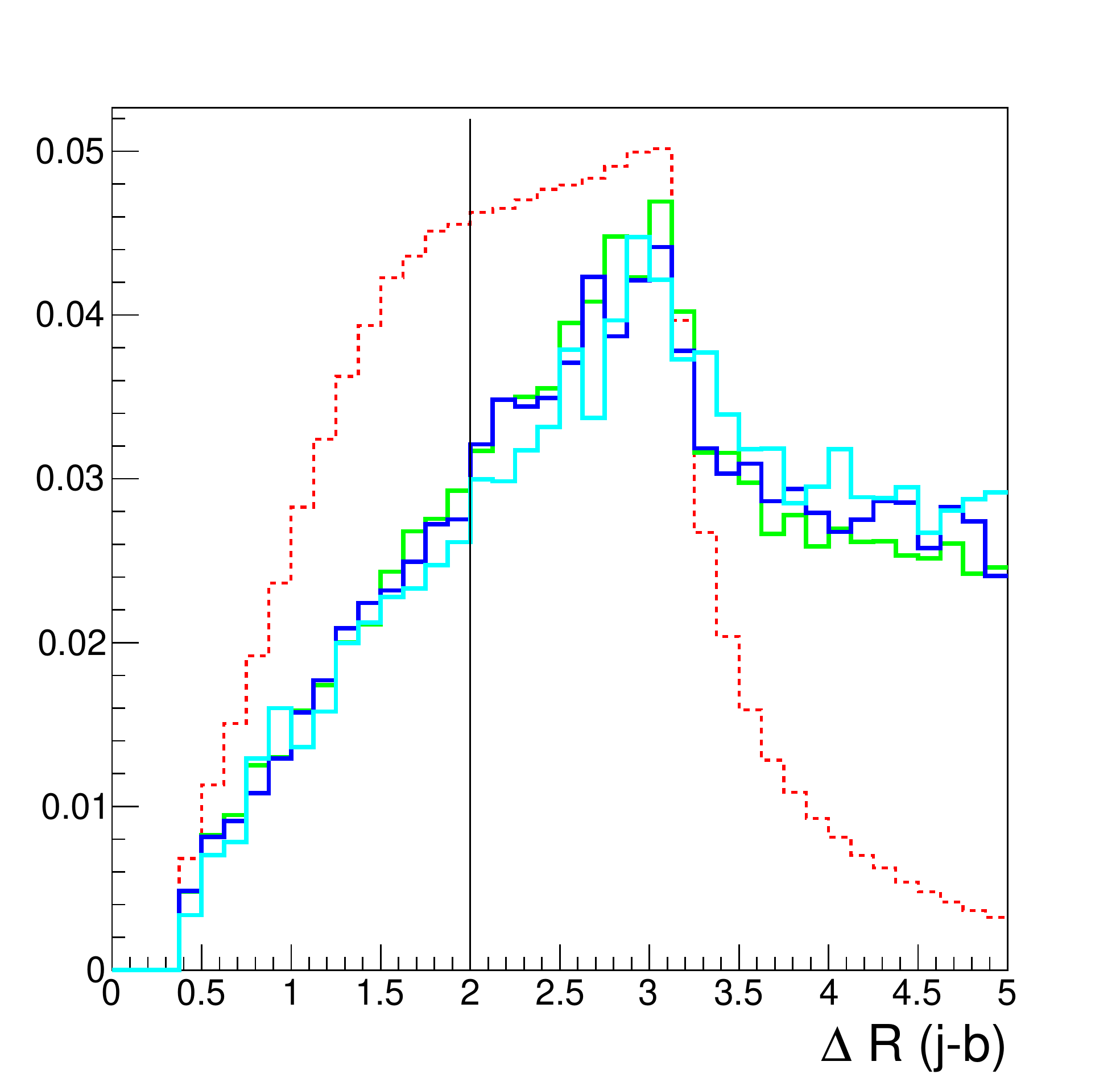}
\caption[]{
\label{deltaR_distribution}
\small
Differential distributions of the $\Delta R$ separations between the tagged $b$-jet and the reconstructed Higgs (left plot) and between the tagged $b$-jet and the tagged light-jet (right plot), after the neutrino and  top
reconstruction, 
for $\sqrt{s} =8\,$TeV. The vertical lines indicate the values of the optimized cuts (\ref{eq:optimized}).
}
\end{center}
\end{figure}

One can see that at this stage the background dominates by far over the signal.
We can however exploit the peculiar kinematics of the signal to perform a set of cuts that reduce the background to 
a much smaller level. 
One of the peculiarities of the signal is the presence of the heavy fermionic resonance among the intermediate final states. Its production requires the exchange of a large amount of energy and leads to very energetic final states. We find very effective applying a cut on the invariant mass of all the $\tilde{T}jb$ particles in the intermediate final state as well as a cut on the transverse momentum of the hardest final jet. We will also apply a cut on the $p_T$ of the reconstructed top and of the reconstructed Higgs. The other important characteristic of the signal topology, as already discussed, is the presence of a very energetic final jet, the tagged light-jet in the $\tilde{T}jb$ final state, which is emitted at very high rapidity. 
We exploit this feature by imposing a cut on the energy and on the rapidity of the tagged light-jet.
Further conditions are imposed on the rapidity of the tagged $b$-jet, which is also emitted at quite high rapidity, and on the $\Delta R$ separations between the tagged $b$-jet and the tagged light-jet, $\Delta R (j-b)$, and between the tagged $b$-jet and the reconstructed Higgs, $\Delta R (h-b)$. These $\Delta R$'s are much larger in the signal than in the background. In particular, 
$\Delta R (b-h)$ is quite small for the main background $t\bar{t}+jets$, where the reconstruction procedure of sec. \ref{sec:reco} tends to select a $W$ (from the hadronic decay of the other non-tagged top) as the Higgs; as a consequence, in the most part of the $t\bar{t}+jets$ events the reconstructed Higgs and the tagged $b$-jet are close to each other, because they both come from the decay of a top. \\
We show in Fig. \ref{pT_distribution} the invariant mass of the system $\tilde{T}jb$, $M_{\tilde{T}jb}$, and the $p_T$ distributions of the hardest jet ($j(1)$), of the reconstructed top and of the reconstructed Higgs. 
In Fig. \ref{MRj_distribution} we show the $|\eta|$ distributions of the tagged light-jet and of the tagged $b$-jet. In Fig. \ref{deltaR_distribution} we report the distributions of the $\Delta R$ separations between the tagged $b$-jet and the tagged light-jet and between the tagged $b$-jet and the reconstructed Higgs.
We show the distributions (normalized to unit area) for the total background and for the signal referred to different $\tilde{T}$ mass values.\\
As expected, the signal from highest $\tilde{T}$ mass values has ever more energetic final particles; especially the distributions of $p_{T}j(1)$ and of $M_{\tilde{T}jb}$ shift on larger values for heavier $m_{\tilde{T}}$.\\
We find a set of optimized cuts that minimizes the integrated luminosity needed for a $5\sigma$ discovery of the signal with $m_{\tilde{T}}=400$ GeV. In the cases of higher $\tilde{T}$ masses we will refine in a second step the cuts on $M_{\tilde{T}jb}$, on $p_{T}j(1)$ and on the top and the Higgs $p_T$'s.
We define the discovery luminosity to be the integrated luminosity for which a goodness-of-fit test of the SM-only hypothesis with Poisson 
distribution gives a  $\text{p-value}=2.85\times 10^{-7}$, which corresponds to a $5\sigma$ significance in the limit of a Gaussian distribution. \\
The optimized cuts (see Figs. \ref{pT_distribution}-\ref{deltaR_distribution}) are:

\begin{equation} \label{eq:optimized}
\begin{aligned}
& M_{\tilde{T}jb}>900 \ \text{GeV} && p_{T}\ j(1)>100\ \text{GeV} \\
&  p_{T}\ top> 90\ \text{GeV} && p_{T}\ h> 90\ \text{GeV} \\
& |\eta_{j}|>2.1 && |\eta_{b}|>0.9 \\
& \Delta R (j-b)>2 && \Delta R (h-b)>1.8 \\
& E(j)>230 \ \text{GeV} && M_{h-j}<70 \ \text{GeV}
\end{aligned}
\end{equation}

The cut on the energy of the tagged light-jet, $E(j)$, is also useful to avoid possible conflicts with the jets from initial state radiation;
$M_{h-j}$ represents the invariant mass of the objects which form the reconstructed Higgs, from which the most energetic jet (among them) is subtracted. The cut on $M_{h-j}$ reduces particularly the backgrounds with more than $5$ jets in the final state.\\
  
After the optimized cuts the background is substantially reduced. At this stage, if we plot the invariant mass of the reconstructed Higgs versus the invariant mass of the reconstructed $\tilde{T}$, for background and signal events, we can clearly distinguish the excess of events in correspondence of the top partner and of the Higgs resonances. 
Once we can recognize these resonances we can also refine the analysis by imposing a cut on $M_{\tilde{T}}$ and on $M_h$. We require $M_h$ to be comprised in the region $[100$ GeV, $150$ GeV] and $M_{\tilde{T}}$ to be in a region of $\pm 2\Gamma(\tilde{T})$ from the $\tilde{T}$ mass value. We also refine the cut on $M_{\tilde{T}jb}$ and on the $p_T$ of the hardest jet, of the top and of the Higgs, according to the values shown on Table \ref{tab:refined}.

%

%

%

%
%

%
\begin{table}
\begin{center}
\begin{tabular}{c|cccc}
  $m_{\tilde{T}}$ (TeV) & $\ M_{\tilde{T}jb}$ & $\ p_T$ $j(1)$ & $\ p_T$ top  & $\ p_T$ h \\[0.15cm]
  \hline 
  \\ 
 0.4 & 0.9 & 0.10 & 0.09 & 0.09\\ [0.1cm]
 0.6 & 1.2 & 0.16 & 0.13 & 0.13\\  [0.1cm]
 0.8 & 1.4 & 0.19 & 0.17 & 0.17\\ [0.1cm]
 1.0 & 1.7 & 0.25 & 0.23 & 0.23\\ [0.1cm]
 1.5 & 2.1 & 0.32 & 0.27 & 0.27\\ [0.1cm]
\end{tabular}
\caption{
\label{tab:refined}
\small 
Refined cuts, in TeV, for different $\tilde{T}$ mass values (first column).
}
\end{center}
\end{table}

The final cross sections for the signal and the main backgrounds, after imposing the optimized cuts of eq.(\ref{eq:optimized}) plus the refined cuts of Table \ref{tab:refined} and the conditions on $M_h$ (to be in the region $[100$ GeV, $150$ GeV]) and on $M_{\tilde{T}}$ (to be in a region of $\pm 2\Gamma(\tilde{T})$ from the $\tilde{T}$ mass value), are reported in Table \ref{tab:cutflow14TeV2} for $\sqrt{s}=8\,$TeV and $\sqrt{s}=14\,$TeV. 
The values of the corresponding discovery luminosity are shown in Table~\ref{tab:Ldisc}.
We do not report the $W+jets$ background, which is negligible at this stage.
For the background, we indicate in parenthesis the one-sigma statistical error on the cross section; 
for the signal, the statistical error is negligible, compared to that of the background, and we do not report it.
Statistical errors on the cross sections are computed by assuming a Poisson distribution for the number of events that pass the cuts \footnote{We calculate the cross section after the application of a cut as
\[
 \sigma= \frac{n}{L}\ ,
\]
where $n$ is the number of simulated events that pass the cut and $L$ is the integrated luminosity reached in the simulation. Given the observed number of events, $n$, the true value of the number of events passing the cut, $\lambda$, follows a Poisson distribution:
\[
 f(\lambda|n)=\frac{\lambda e^{-\lambda}}{n!}\ .
\]
The variance associated with $\lambda$ is $Var[\lambda]=n+1$. We thus associate to the cross section a variance:
\[
 Var[\sigma]=\frac{n+1}{L^2} \ .
\]
\noindent
When we sum over different cross section values, the error is summed in quadrature.}.
In order to obtain a conservative estimate of the discovery luminosity, we consider the central value plus one-sigma as the value of the background cross section.

\begin{table}
{\small
\begin{tabular}{|c|c|c|c|c|}
\hline 
 & & &  &\\[-0.1cm]
\multicolumn{1}{|l|}{\textsf{LHC  $\mathsf{8\,}$TeV}}& \multicolumn{1}{l|}{Signal} & \multicolumn{1}{l|}{$WWbb+jets$} & \multicolumn{1}{l|}{$Wbb+jets$} & \multicolumn{1}{l|}{TOT Bckg}\\[0.1cm]
\hline
& & &  & \\[-0.3cm]
$m_{\tilde{T}} = 0.4\,$TeV & 3.67 & 3.0(1) & 0.10(1) & 3.1(1) \\[0.15cm]
$m_{\tilde{T}} = 0.6\,$TeV & 0.865 & 0.22(3) & 0.033(5) & 0.25(8) \\[0.15cm]
$m_{\tilde{T}} = 0.8\,$TeV & 0.270 & 0.03(1) & 0.015(4) & 0.04(1) \\[0.15cm]
$m_{\tilde{T}} = 1.0\,$TeV & 0.060 & 0.007(7) & 0.006(2) & 0.013(8) \\[0.15cm]
\hline
\end{tabular}
\\[0.4cm]
\begin{tabular}{|c|c|c|c|c|}
\hline 
 & & &  &\\[-0.1cm]
\multicolumn{1}{|l|}{\textsf{LHC  $\mathsf{14\,}$TeV}}& \multicolumn{1}{l|}{Signal} & \multicolumn{1}{l|}{$WWbb+jets$} & \multicolumn{1}{l|}{$Wbb+jets$} & \multicolumn{1}{l|}{TOT Bckg}\\[0.1cm]
\hline
& & &  & \\[-0.3cm]
$m_{\tilde{T}} = 0.4\,$TeV & 20.5 & 23(1) & 0.79(4) & 25(1) \\[0.15cm]
$m_{\tilde{T}} = 0.6\,$TeV & 6.46 & 2.3(3) & 0.32(3) & 2.7(3) \\[0.15cm]
$m_{\tilde{T}} = 0.8\,$TeV & 2.44 & 0.3(1) & 0.15(2) & 0.4(1) \\[0.15cm]
$m_{\tilde{T}} = 1.0\,$TeV & 0.721 & 0.02(3) & 0.06(1) & 0.08(3) \\[0.15cm]
$m_{\tilde{T}} = 1.5\,$TeV & 0.066 & 0.00(1) & 0.004(4) & 0.00(1) \\[0.15cm]
\hline
\end{tabular}
\caption{
\label{tab:cutflow14TeV2}
\small 
Cross sections, in fb, at $\sqrt{s}=8\,$TeV (upper table) and at $\sqrt{s}=14\,$TeV (lower table) for the signal (with $\lambda_{\tilde{T}}=3$) and the main backgrounds after imposing the optimized cuts of eq. (\ref{eq:optimized}) plus the refined cuts of Table \ref{tab:refined} and the restrictions: $M_h \in $ [100 GeV, 150 GeV], $M_{\tilde{T}}\in m_{\tilde{T}}\pm 2\Gamma(\tilde{T})$. For the background, we indicate in parenthesis the one-sigma statistical error on the cross section. 
}}
\end{table}

%

\begin{table}[h]
\begin{center}
%
\vspace{0.5cm}
\begin{tabular}{r|ccccc}
\multirow{2}{*}{\textsf{LHC $\mathsf{\sqrt{s} = 8\,}$TeV}} & \multicolumn{3}{c}{$m_{\tilde{T}} [\text{TeV}]$} & \\[0.15cm]
  & 0.4 & 0.6 & 0.8 & 1.0\\[0.1cm]
\cline{1-5}
& & & & \\[-0.3cm]
$L_{disc} \,[\text{fb}^{-1}]$ & 7.8 & 17 & 40 & 260
\\
\multicolumn{4}{c}{}
\\[0.4cm]
\multirow{2}{*}{\textsf{LHC $\mathsf{\sqrt{s} = 14\,}$TeV}} & \multicolumn{5}{c}{$m_{\tilde{T}} [\text{TeV}]$} \\[0.15cm]
 & 0.4 & 0.6 & 0.8 & 1.0 & 1.5 \\[0.1cm]
\hline
& & & & &\\[-0.3cm]
$L_{disc} \,[\text{fb}^{-1}]$ & 2.0 & 2.9 & 4.7 & 13 & 150
\end{tabular}
\vspace{0.5cm}
\caption{
\label{tab:Ldisc}
\small 
Value of the integrated luminosity required for a $5\sigma$ discovery, 
for $\sqrt{s}=8\,$TeV (upper panel) and $\sqrt{s}=14\,$TeV (lower panel).}
\end{center}
\end{table}

\subsection{Discovery reach on the $\mathbf{(m_{\tilde{T}},\lambda_{\tilde{T}})}$ plane}

\begin{figure}
\includegraphics[width=0.5\textwidth,clip,angle=0]{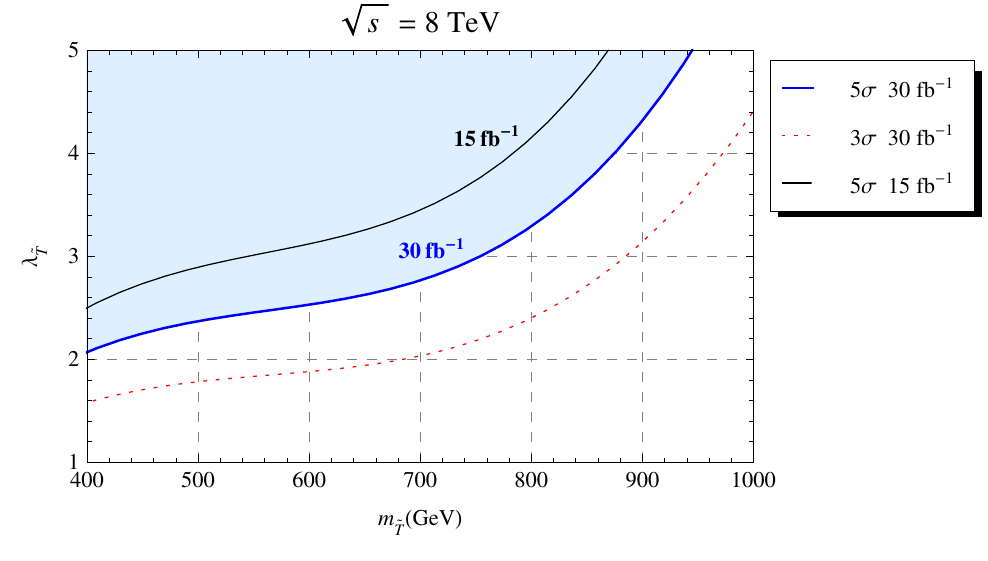}
\includegraphics[width=0.49\textwidth,clip,angle=0]{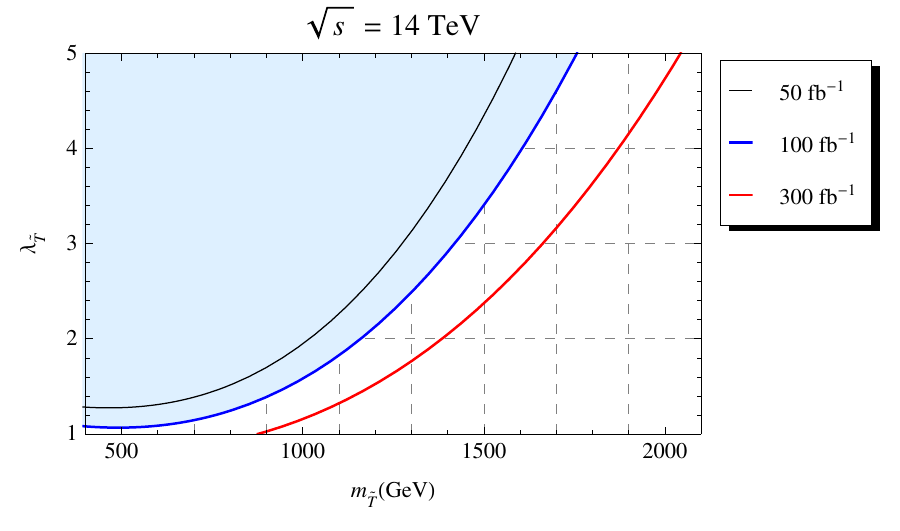}
\caption[]{
\label{fig:reach}
\small
LHC discovery reach in the plane $(m_{\tilde{T}},\lambda_{\tilde{T}})$ for the signal $pp\to (\tilde{T}\to (h\to bb)t)b+X$. Upper plot: LHC at $\sqrt{s}=8\,$TeV; the blue area shows the region where a discovery of the signal is possible at $5\sigma$ with $L = 30\,\text{fb}^{-1}$; the black continous  (dotted) curve defines the region of a $5\sigma$ ($3\sigma$) discovery with $15$ ($30$) fb$^{-1}$. Lower plot: LHC at $\sqrt{s}=14\,$TeV; the blue area shows the $5\sigma$ discovery region with $L = 100\,\text{fb}^{-1}$; the black (red) curve defines the region of a $5\sigma$ discovery with $50$ ($300$) fb$^{-1}$.
}
\end{figure}

All the numbers shown in Tables~\ref{tab:cutflow14TeV2} and~\ref{tab:Ldisc} hold for a fixed coupling $\lambda_{\tilde{T}}=3$. 
As also discussed in Ref. \cite{Vignaroli:2012Bs}, it is particularly interesting to study the dependence of our results on the coupling $\lambda_{\tilde{T}}$; this, indeed, could give us 
an estimate of the LHC sensitivity to measure the Higgs (and electro-weak bosons) coupling to the top partner, for different masses of this latter, and, consequently, to obtain information on the mechanism behind the EWSB. 
We can generalize our results to different $\lambda_{\tilde{T}}$ values, by simply considering that the production cross section scales with $\lambda^2_{\tilde{T}}$. 
It is thus possible to estimate how the LHC discovery reach
varies with $\lambda_{\tilde{T}}$ by simply rescaling the numbers in Table \ref{tab:cutflow14TeV2} to take into account the change in the production cross section.  
The result is reported in Fig.~\ref{fig:reach}. The two plots show the 
region in the plane $(m_{\tilde{T}},\lambda_{\tilde{T}})$ where a $5\sigma$ discovery is possible for the LHC at $\sqrt{s}=8\,$TeV with $L = 30\,\text{fb}^{-1}$ and $L = 15\,\text{fb}^{-1}$
(upper plot), and at $\sqrt{s}=14\,$TeV with $L = 50,\ 100,\ 300\ \text{fb}^{-1}$ (lower plot).

\section{Discussion}\label{sec:conclusions}

Our results are summarized by Fig. \ref{fig:reach}. They show that, for a reference value $\lambda_{\tilde{T}}=3$ of the Higgs coupling to the top partner $\tilde{T}$, the 8 TeV LHC with $30$ fb$^{-1}$ can discover the singly-produced top partner in the channel $pp\to (\tilde{T}\to (h\to bb)t)b+X$ if the top partner has a mass up to $760$ GeV (while an observation is possible for $m_{\tilde{T}}\lesssim 890$ GeV). If the LHC and Tevatron excesses near $125$ GeV are really due to a composite Higgs, naturalness arguments demand top partners below $\sim 1$ TeV.  Our results highlight thus that the 8 TeV LHC already has a large sensitivity on probing the composite Higgs hypothesis.\\
The LHC reach 
is even wider at $\sqrt{s}=14$ TeV.  With $\lambda_{\tilde{T}}=3$, the LHC with $100$ fb$^{-1}$ can observe (at 5$\sigma$) a Higgs from a top partner decay for masses of this latter up to $\simeq 1450$ GeV; in the case the top partner was as light as $\simeq 500$ GeV, the 14 TeV LHC would be sensitive to the measure of the $\lambda_{\tilde{T}}$ coupling in basically the full range $\lambda_{\tilde{T}}>1$ predicted by the theory.\\
The single production of the top partner is thus a very promising channel to observe the Higgs and to test its possible composite nature. It proves to be a promising channel for the discovery of the top partner itself. One could also consider that, since $BR(\tilde{T}\to ht)\simeq BR(\tilde{T}\to Zt)$, results similar to those obtained in this analysis are expected from the $(\tilde{T} \to (Z\to hadrons)t)b+X$ channel if one adopts the strategy outlined here, with a variation in the cut on the Higgs ($Z$) invariant mass. Our results for the discovery of top partners are comparable to those in \cite{Mrazek:2009yu}, where both the single and double productions of bottom and exotic partners are analyzed, and competitive with those in Refs. \cite{Bini, Santiago}, which have considered the production of heavy fermions in association with their SM partners from the decay of a heavy gluon. \\
Finally, we point out that the top partners considered in this analysis 
are a very general prediction of composite Higgs models and theories with a warped extra dimension; the argument that they are expected to be lighter than $1$ TeV is robust, since it is related to naturalness.  
This is not the case for exotic and for (some types of) bottom partners, that could either be absent or heavier than $1$ TeV.




\end{document}